\pdfoutput=1
\documentclass[11pt]{article}

\usepackage[final]{acl}

\usepackage{times}
\usepackage{latexsym}

\usepackage[T1]{fontenc}
\usepackage[utf8]{inputenc}

\usepackage{microtype}

\usepackage{inconsolata}

\usepackage{graphicx}

\title{Large Language Model Use Impact Locus of Control}

\author{Jenny Xiyu Fu \\ 
  Cornell University \\
  Ithaca, New York, USA \\
  \texttt{xf89@cornell.edu} \\\And
  Brennan Antone \\
  Cornell University \\
  Ithaca, New York, USA \\
  \texttt{brennan@cornell.edu} \\\And
  Kowe Kadoma \\
  Cornell University \\
  Ithaca, New York, USA \\
  \texttt{kk696@cornell.edu} \\\And
  Malte Jung \\
  Cornell University \\
  Ithaca, New York, USA \\
  \texttt{mfj28@cornell.edu} \\}

\begin{document}
\maketitle
\begin{abstract}
As AI tools increasingly shape how we write, they may also quietly reshape how we perceive ourselves. This paper explores the psychological impact of co-writing with AI on people's locus of control. Through an empirical study with 462 participants, we found that employment status plays a critical role in shaping users' reliance on AI and their locus of control. Current results demonstrated that employed participants displayed higher reliance on AI and a shift toward internal control, while unemployed users tended to experience a reduction in personal agency. Through quantitative results and qualitative observations, this study opens a broader conversation about AI's role in shaping personal agency and identity.
\end{abstract}

\section{Introduction}
For centuries, from ink on fabric to text on screens, writing has always been a powerful tool for self-expression, providing a space for people to define, reflect, and communicate their sense of self to the world~\cite{channa2017letter}. While the mediums have changed over time, the act of writing remains deeply personal— a way to explore and communicate our identities~\cite{park2013writing, ivanivc1998writing, pennebaker2007expressive}. Now, generative AI steps into this long tradition, adding a layer of collaboration in the creative process. While the benefits of AI are clear (e.g., reduced mental strain, enhanced productivity, and stronger persuasive writing)~\cite{Zhang-2023-productivity}, its influence extends beyond mere assistance, and the nuanced psychological impacts of AI use in writing remain underexplored. 

As AI systems become more involved in the writing process, they take on an active role in shaping people's narratives, which may lead to a diminished sense of ownership and agency over these narratives~\cite{kadoma2024role,mieczkowski2021ai}. Thus, questions arise about how these technologies might reshape this age-old relationship. How might the use of AI in writing affect one's sense of self? Could this involvement with AI alter our personal agency over the narratives we produce? We recognized the potential risks of AI in diminishing people's ownership of the narratives they are crafting, undermining their sense of self. 

Locus of control, which refers to the extent to which individuals believe they have power over the outcomes of events in their lives, is a particularly effective measure of self-agency because it directly assesses whether individuals feel responsible for their actions and outcomes~\cite{rotter1966generalized}. This concept has been used as a principal of interface design to evaluate users' agency with the system from the early research in HCI, when Ben Shneiderman proposed the concept of universal usability to empower individuals in 2000~\cite{shneiderman2000universal}. In contexts where individuals need to craft their personal or professional identity, locus of control is critical to evaluate as AI's involvement can become an external force for the user, blurring the lines between self-driven expression and system-driven narratives.

People with an internal locus of control tend to perceive themselves as active agents in shaping their environment, leading to a stronger sense of agency and ownership over personal narratives. This internal perspective fosters a stronger, more coherent sense of self, as individuals believe their efforts and decisions directly influence their lives. In contrast, those with an external locus of control may feel their lives are shaped by outside forces, which can lead to feelings of helplessness or detachment from their own identity~\cite{gangai2016association,strauser2002relationship,hiroto1974locus}. When receiving AI suggestion in writing, individuals experience a stronger internal locus of control tend to have a more empowered view on their narrative construction, while individuals who feel a shift towards external forces might felt their self-perceptions is weaken by the AI system.

In professional settings, platforms such as LinkedIn and Indeed provide valuable case studies for exploring the psychological effects of AI co-writing. These platforms prompt users to create profiles that summarize their professional experiences and expertise, with AI increasingly playing a role in generating suggestions for profile content~\cite{cohen2023unlock,batty2023attract}. As AI takes on a greater role in crafting these narratives, it introduces the risk of blurring the boundaries between personal ownership and agency. Given that AI-generated suggestions tend to lean positive~\cite{hohenstein2023artificial,10.1145/3613904.3641955}, and considering prior findings on social media's reflective role in identity formation~\cite{gonzales2008identity}, there's a potential risk that co-writing with AI system might undermine users' sense of self.

Expanding on the critical role of user profiles on professional platforms, this paper explores how AI co-writing can potentially strengthen or weaken users' sense of control, considering the diverse requirements of job seekers and employed individuals. As AI systems take on more active roles in assisting with communication tasks—ranging from auto-completing sentences to drafting paragraphs—their deeper implications for professional identity and personal well-being remain complex. We aim to highlight the varied effects of AI prompts across different demographics, thereby advancing our understanding of the possible pitfalls of personalized AI. We demonstrate that well-intentional prompt engineering, if not critically evaluated directly with end-users, may inadvertently compromise self control. Our observations - on how unemployed job seekers, a low-status group~\cite{smith2012status}, may receive less support from wellbeing AI - form the basis of this discourse.

\section{Background}
Before the era of generative AI, professional social platform like LinkedIn and Indeed have been a long standing artifact in the study of self presentation by HCI and communication scholars. On these platforms, users make deliberate choices about their professional self-presentation~\cite{domahidi2022you}; and research underscores the significance of user profiles: comprehensive profiles are more likely to secure employment and receive promotions~\cite{groysberg2023linkedin,davis2020networking}. As interest in using AI in professional settings grows, many platforms are adding features like auto-complete and AI writing assistance. LinkedIn, for example, has introduced its in-house AI system to write profile summaries and headlines, and Indeed rolled out AI job description generator~\cite{cohen2023unlock,batty2023attract}. AI-mediated writing complicates self-presentation, with viewers facing challenges of recognizing authenticity and professionalism from profiles~\cite{mieczkowski2022ai,jakesch2019ai}, and creators mixing personal expressions with AI-improved enhancement\cite{jakesch-etal-2023-cowriting,kadoma2024role,10.1145/3613904.3641955}.

The creation of coherence writing requires the person to register and organized their thoughts, and concretly articulate a storyline. The act of articulating one's career journey can be not merely administrative but profoundly empowering~\cite{ibarra2004working}. One related field on this topic is the psychology of storytelling, which highlights storytelling as a tool for identity formation. Through narrative construction, individuals reflect on their experiences, evaluate their accomplishments, and integrate these events into their sense of self. In the context of professional self-presentation, the attempts of creating coherent narratives allows a deeper self-reflection, reinforcing self-concepts. By narrating their career paths, individuals not only convey their qualifications to employers but also make sense of their own professional journey. 

This articulation prompts people to recognize both their strengths and qualifications as well as their perceived shortcomings. While identifying one's achievements can be empowering, confronting gaps may lead to harmful negative self-thoughts. For individuals with lower self-esteem or in vulnerable periods of unemployment or career transition, this self-assessment process can exacerbate feelings of inadequacy. Research shows that people prone to depression often attribute failures to internal factors (blaming themselves) while externalizing successes (crediting luck or others)~\cite{Kuiper1978DepressionAC}. As a result, the act of professional self-presentation, rather than boosting self-esteem, may lead to a spiral of negative thoughts and increased anxiety. As AI systems become more integrated into the process of professional self-presentation, it is important to understand how people respond to the AI suggestions when crafting their professional narratives. Current study is motivated to investigate how users engage with AI auto suggestions when writing their profiles, assessing the impact of these interactions on their sense of agency through the lens of locus of control.

\section{Methods}
\subsection{Participant Recruitment}
We recruited 462 participants through Prolific, ensuring a gender-balanced cohort of English-speaking adults based in the US. Ranging from 18 to 85 years old, the average age of participants was 39.4 years old (SD= 13.97). Participants were predominantly White (68.4\%), followed by Asian (9.3\%), Mixed (8.9\%), Black (7.8\%), and Other (5.6\%) racial groups. Employment statuses varied, with 50\% full-time employees, 18.2\% part-time workers, 16.9\% unemployed job seekers, and 14.9\% in other categories such as homemakers, retirees, or those about to start new jobs.

\subsection{Study Procedure}
Participants were briefed on a writing task limited to 5-7 sentences, with the average completion time for the survey being around 9 minutes. Upon consenting to participate, participants first completed a personality assessment on their current perceptions of locus of control. The writing task started with participants submitting their current LinkedIn profile, removing any personal identifiers. In the absence of a LinkedIn summary, participants were instructed to note, "I currently do not have a summary." This was followed by a motivational primer underscoring the importance of a polished professional online presence, citing a Harvard Business Review study that linked detailed online profiles to potential salary increases of 2-8\%~\cite{groysberg2023linkedin}. The primer aimed to engage participants fully and enhance their focus on the task. Subsequently, the main writing task involved crafting a LinkedIn profile summary. Participants were randomly assigned to one of four conditions: no AI, default AI, AI emphasizing external locus of control, and AI emphasizing internal locus of control. After completing the writing task, they revisited the initial assessments on their perceptions of locus of control. The research design was approved by the University IRB.

\subsection{The Writing System}
\begin{figure}[t]
\includegraphics[width=\columnwidth]{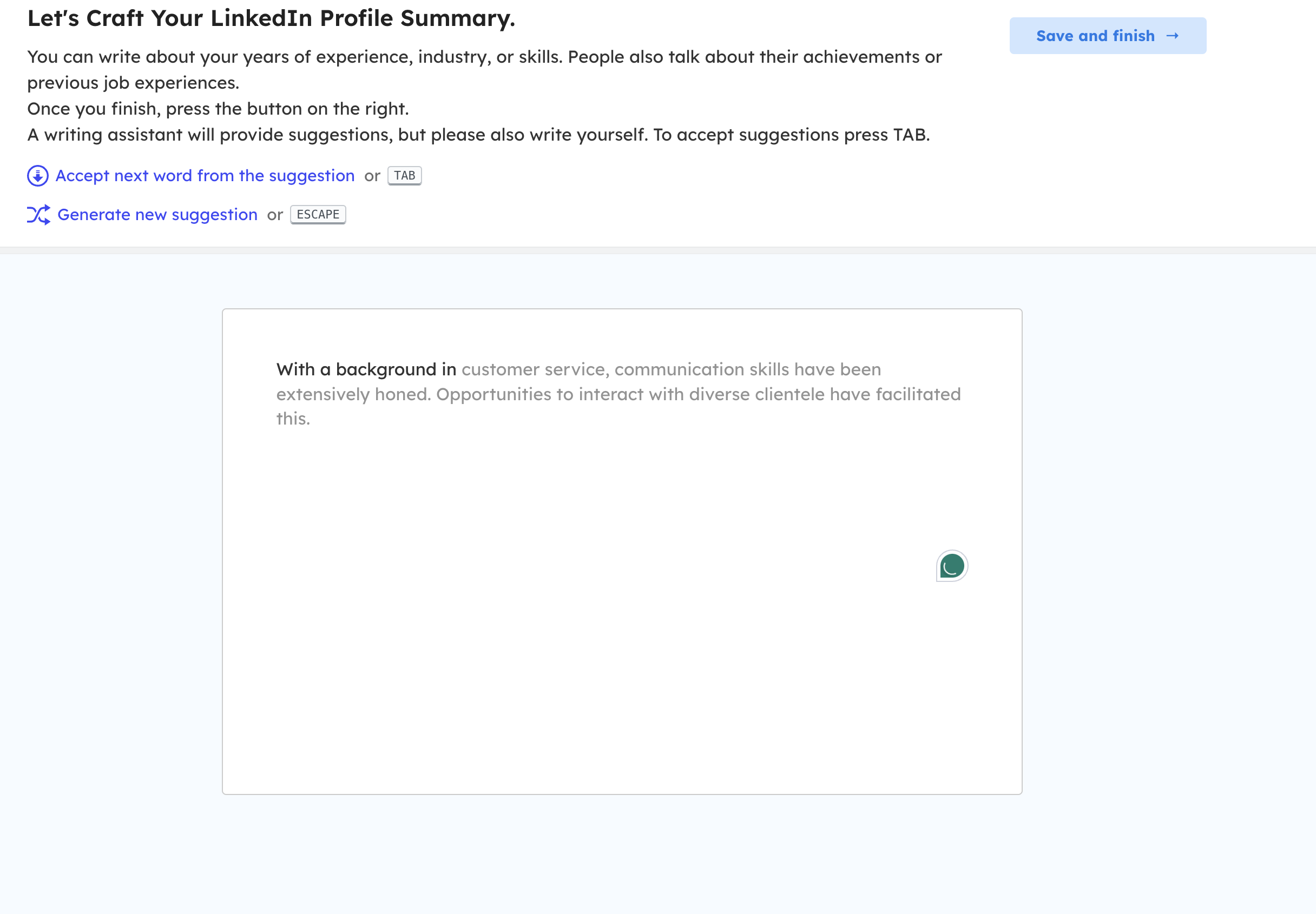}
\caption{\textbf{Screenshot of the writing task.} Instructions are displayed at the top panel: Press `tab' to accept a suggestion or `esc' to request a new one. Written text is shown in black, while suggestions are in gray.
}
\label{fig:writing_interface}
\end{figure}

We utilized the GPT-4 API to generate prompts tailored to each experimental condition, aiming to highlight narratives that aligned with the specified locus of control orientation. An illustrative demonstration of the interface layout is depicted in Figure~\ref{fig:writing_interface}. The foundation for the writing interface was adapted from the codebases of prior research~\cite{Jakesch-hueristics-2023}. The mechanism operates by whenever a user stops typing, the system presents recommendations of no more than 2 sentences, each sentence less than 10 words. The participant has the option to adopt the suggestion by hitting the tab or the right arrow key, or to dismiss it by resuming typing. Hitting the tab key incorporates the next word from the suggestion, and the user can press it repeatedly to accept the entire recommendation. In the external locus of control condition, the AI was directed to adopt a formal, factual tone, using third-person passive voice and a neutral tone. Conversely, in the internal locus of control condition, the AI was prompted to create conversational, storytelling narratives in the first-person active voice, fostering a positivist tone.

\subsection{Survey Measurements}
Our analysis focuses on examining how different writing conditions, such as no AI assistance, internal or external locus of control emphasis, and the default GPT-4 model, impact participants, especially those seeking jobs or currently employed. Three measurements are used to capture the effects: AI reliance, writing time, and locus of control. 

\textbf{Writing Time}: The total time participants spent crafting their profile summary, this is measured by the total amount of time participants spent on the writing interface. 

\textbf{AI Reliance}: Introduced by Kadoma et al.~\cite{kadoma2024role}, AI reliance is the fraction of AI-written characters in the entire text. This is used to measure the contribution of AI in the final written paragraph. Ranging from 0 to 1, the score with 1 means the assistant wrote the entire text (reflecting complete reliance on the assistant) and 0 means the human wrote the entire text (no reliance). For example, in the control condition, since there was no AI introduced, the AI reliance will be 0. 

\textbf{Locus of control}: The study employed the IE-4 scale to measure locus of control~\cite{niessen2022internal}. The scale includes four items to capture the degree of one's belief on their power and control over events in their lives: two reflecting an internal locus (1- "I'm my own boss"; 2- "If I work hard, I will succeed") and two indicating an external locus (3- "Whether at work or in my private life: What I do is mainly determined by others"; 4- "Fate often gets in the way of my plans"). Participants were asked to rate on a scale from 1 to 5, 1 means does not apply all, and 5 means applies completely. To assess the impact of writing with AI assistance on participants' locus of control (LoC), we analyzed the overall mean change across all four items in the LoC measurements from before to after the intervention (Post - Pre). When the value of differences is negative, it indicates the shift towards external locus of control; sequentially, when the value is positive, it indicates a shift towards internal locus of control.

\section{Results}

\subsection{Writing Time}
Across all conditions, the average writing time was 4.89 minutes (SD = 4.46), with employed participants generally spending less time writing (M = 4.77, SD = 4.55) compared to those not employed (M = 5.17, SD = 4.20). Examining each condition, participants writing without AI assistance took the longest time to craft their profiles, averaging 5.54 minutes. This was closely followed by those using external AI, who took 5.48 minutes. Participants with the default AI setting took 4.36 minutes, while the group using internal AI demonstrated the shortest average writing time at 4.04 minutes.

\begin{figure}[t]
\includegraphics[width=\columnwidth]{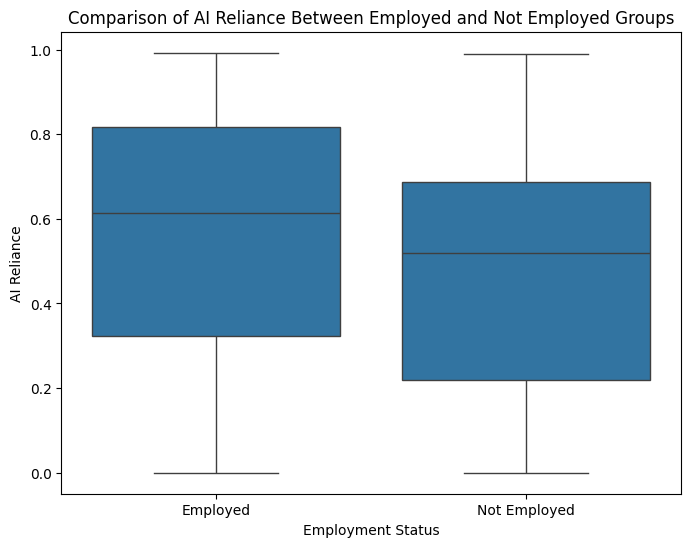}
\caption{\textbf{AI Reliance by Employment Status.} This boxplot illustrates the distribution of AI reliance scores among different employment groups.}
\label{fig:Reliance_Employment}
\end{figure}

\begin{figure}[t]
\includegraphics[width=\columnwidth]{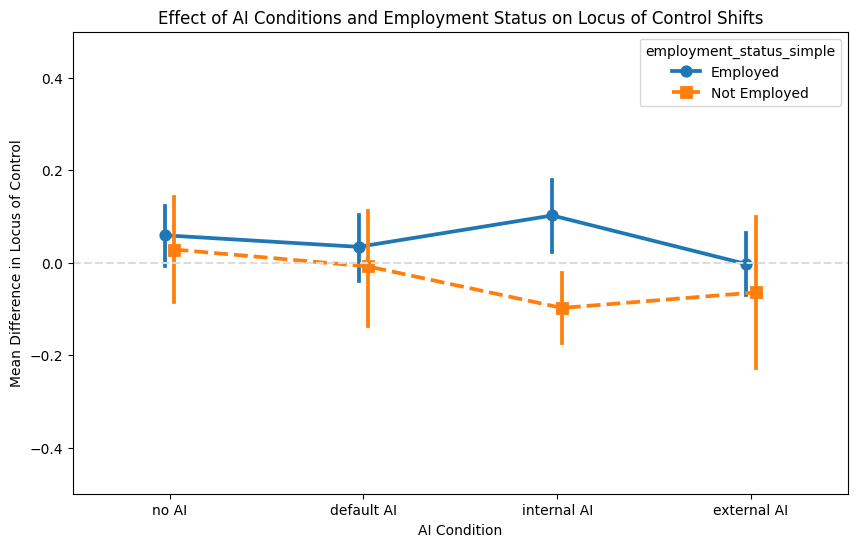}
\caption{\textbf{Locus of Control Shifts Across AI Conditions and Employment Status.} This graph displays the mean differences in locus of control among employed and unemployed participants, sorted by AI writing conditions.}
\label{fig:LoC_Employment}
\end{figure}

\subsection{AI Reliance}
We observed a nuanced relationship between employment status and AI reliance, highlighting how employment circumstances might influence one's experiences when writing with AI. Specifically, employed individuals demonstrated a greater level of reliance on the AI system (M = 0.554, SD = 0.308) compared to their not-employed counterparts who showed a lower average AI reliance (M = 0.469, SD = 0.293), shown in Figure ~\ref{fig:Reliance_Employment}. This significant difference between the groups (t = 2.28, p = 0.023) indicates that employment status shapes the dynamics of AI use. When dissecting the data further by the AI writing style, distinct patterns emerged. Participants relied more on the default AI (M = 0.604, SD = 0.273), followed by the internal AI(M = 0.561, SD = 0.298), and relied on the AI model the least when they were writing with the external AI(M = 0.443, SD = 0.320). The differences are statistically significant (F-value=9.248, p<0.001), particularly between the 'Default' and 'External' groups, and between the 'External' and 'Internal' groups, indicating that the context of AI reliance—whether default, external, or internal—significantly influences the perceived mutuality in AI interactions.

\subsection{Locus of Control}
The findings highlight an interaction between employment status and AI writing style: depending on their current employment status, writing with internal AI can either enhance or diminish an individual's sense of professional agency. Figure ~\ref{fig:LoC_Employment} illustrates the change of participants' locus of control across all four conditions and the differential impact of employment status on these shifts. Current results showed, employed individuals reported a significant shift towards internal sense of control compared with the unemployed individuals who had a shift toward external sense of control (t = 2.49, p = 0.014, Cohen's d = 0.58). 

Overall, participants showed a slight increase in the locus of control (M = 0.027, SD = 0.35), suggesting a shift toward an internal LoC. Specifically, employed participants experienced an increase in LoC (M = 0.048, SD=0.33), whereas not employed participants observed a decrease (M = -0.03, SD=0.39). This difference is statistically significant (t= 2.12, p=0.034). This significant result suggests that employment status has a notable impact on how individuals' LoC changes in response to completing the professional narrative creation. Delving into the data by treatment groups, the Internal AI group exhibited a slight increase in LoC (M = 0.06, SD = 0.35), in contrast to the External AI group, which showed a minor decrease (M = -0.02, SD = 0.39). There is no significant differences between the no AI condition and the other three AI conditions. Since our current study is an exploratory analysis, we are interested in the effect of writing with which kind of AI, which is the first step in designing effective AI interventions. 

\section{Qualitative Observation}
To understand how participants in different employment statuses write differently, we collected the writing interaction log data during the experiments and presented the following qualitative observations. We manually processed the participants' written texts and highlighted emerging patterns. The approach is not intended as a systematic qualitative analysis but rather an interpretive observation of the written texts, and the purpose of recognizing the observations is to add depth to the quantitative findings by offering an additional dimension of participant differences. We begin by reviewing the conversational writing dynamic participants exhibit with the AI system. This is followed by a discussion on different strategies participants had when they recognized a tone switch from the AI suggestions. We observed that comparing with unemployed participants, employed participants demonstrate a stronger control of self during co-writing, and their written outcome has a stronger sense of self. 

At the beginning of the writing, both employed and unemployed participants revealed patterns of engaging in a conversational dynamic with the system, adjusting the AI's suggestions to better align with their own writing style. Zooming into the interaction replay, we observed that participants responded differently to the AI's suggestions. Besides the system design's affordances allowing participants to accept or regenerate the suggestion, many chose to rephrase or modify the AI-generated content themselves. For the employed participants, they showed more sense of control in creating coherence in writing by repetitively regenerating the AI suggestions. Upon receiving AI suggestions, employed participants would pause to evaluate and regenerate responses. If the suggestions remained unsatisfactory, they often bypassed the AI, wrote on their own terms for the current sentence, and repeated the AI generation for the following sentence. In contrast, unemployed participants reacted more passively with undesirable AI suggestions. Rather than rejecting or rephrasing the AI's outputs, they were more likely to accept most of the AI suggestions in the rest of the writing. 

Participants' final writing output also demonstrated the differences in control of self. In employed participants' writing, more sentences began with personal pronouns like "I," emphasizing their active role in shaping the narrative. 
Here is an example of an employed participant's writing: ``I have worked in social services for the past ten years. I am trained as a social worker, having completed my Masters in Social Work from the University of Utah in 2019. External influences have greatly shaped the direction of this path. Shortly thereafter, I stepped outside of this role to raise my daughter. I wanted to try something new having just recently reentered the workforce and I am currently working as an independent contractor doing housekeeping and providing transcription services.'' In this example, the participant selectively accepted AI-generated suggestions, ensuring that the writing retained a personal and coherent narrative. 

Comparing with employed participants who demonstrated a higher sense of control by preserving their own voice, unemployed participants exhibited a more passive approach in co-writing. After the first sentence, they tended to accept most AI suggestions without much modification. Their final written text thus shifted towards a more factual, impersonal tone, which is more similar to the prompted style for the External AI condition design. Here is an example of an unemployed participant's writing: 
``I have many years of experience in team leadership and project management, molded by constant interaction with diverse teams. The success of these teams was greatly driven by external collaborations. External factors have significantly influenced my career trajectory. These factors included market trends and economic climates. The unpredictable nature of these elements often necessitated adaptability.'' 

The qualitative interaction offers a dimension in understanding the observed statistical analysis, where there is a significant shift toward external locus of control in unemployed participants. Despite the current results showing that unemployed participants have a lower AI reliance across all conditions, the additional cognitive effort required to process and reject these suggestions—especially when connected to their professional identity—may still increase their cognitive load. For unemployed participants, the increased cognitive load introduced by integrating these suggestions may overwhelm their capacity to assert personal control. When writing with an External AI, the act of resisting AI suggestions may itself be taxing, contributing to feelings of diminished personal agency, resulting in the observed shift toward an external locus of control.

\section{Discussion}
Our findings extend the existing research on the user experiences of "writing with AI" by examining the impact of AI-assisted writing and locus of control. The relationship between employment status and AI reliance suggests that user experiences with AI are deeply influenced by their personal and professional circumstances.

We found that employment status significantly influences how individuals engage with AI in writing tasks, with unemployed participants showing a lower reliance on the AI assistant. One potential explanation for our observations lies in the cognitive experiences of participants. For the unemployed individuals, engaging with AI-assisted writing might have evoked a heightened sense of uncertainty regarding their professional identity. Since unemployed individuals can belong to the low-status group, they experience more cognitive load and stress when presenting themselves. They have been shown in prior work to experience greater cognitive challenges in the job search process due to this diminished sense of personal status~\cite{smith2012status,menon2014identities}. Thus, this group could perceive the AI's input as unhelpful and also as an external influence on their self-presentation, possibly reinforcing a perception that their professional success is more contingent on external factors than their own abilities.  

It could also be possible that the unemployed participants might experience a form of cognitive dissonance when using AI to enhance their professional profiles. While the AI aims to help them present themselves more effectively, its assistance might also subtly remind them of their current employment status, potentially leading to a sense of reliance on external tools for personal advancement. This could diminish their sense of personal agency, inadvertently shifting their locus of control further outward. Therefore, it is especially concerning given that the benefits of some of the AI systems were primarily observed among employed individuals.

In contrast, employed individuals might interact with the AI from a secure position, viewing it as a supplementary tool that enhances their existing capabilities rather than as a crutch. This perspective could reinforce their internal locus of control, affirming their belief in their ability to shape their professional narratives and outcomes directly. For this group, AI assistance in writing might serve as an empowering tool, validating their competencies and amplifying their sense of autonomy and control over their career progression.

While these are plausible explanations for our findings, identifying causal mechanisms was beyond the scope of this study. Our goal is to start a conversation by describing the interactions between "writing with AI" and individual psychological states, especially how AI interventions can be tailored to foster a positive sense of control and autonomy in users' professional lives, regardless of their current employment situation. To design an inclusive human-AI interaction, we need to consider the diverse psychological needs of users. The diverged shift of the locus of control should raise our awareness of how well-intentional design might disproportionately impact vulnerable communities, and addressing these disparities is important in ensuring that the AI system is supporting growth for all users.

\subsection{Design Recommendation}
Based on the current findings, we propose several design recommendations related to AI's effect on users' locus of control that academic researchers and field practitioners can explore further:
\begin{itemize}
    \item \textbf{Adaptive Moment-to-Moment Interaction}: To alleviate the cognitive burden observed in vulnerable populations, AI interfaces should prioritize simplicity and clarity in decision-making. The system can implement dynamic assistance levels that adjust based on user engagement, offering more or fewer suggestions depending on how actively users interact with the AI. 
    \item \textbf{User Context-Aware Assistance}: Based on our findings that psychological responses to AI writing assistance vary significantly by user context, we recommend systems that adapt to users' specific circumstances and needs. Interfaces should detect patterns of engagement that might indicate reduced agency or heightened cognitive load, then adjust suggestion style and frequency accordingly. For unemployed users who showed reduced agency, systems can offer fewer but more affirming suggestions that emphasize user control. For employed users who demonstrated higher AI reliance, systems can provide more efficiency-focused, comprehensive assistance.
    \item \textbf{Encourage Active Rewriting}: Additionally, the platform can offer supporting features like active reflective writing, encouraging users to modify the co-writing at their own pace. Incorporating the option to engage in wellness-related writing could build emotional resilience and reinforce users' sense of ownership over their narrative.
\end{itemize}

\subsection{Limitations and Future Work}
While our study provides insights into how employment status influences interactions with AI writing assistants, several limitations should be noted. Finding a job is an iterative process, and updating a professional narrative takes time. Currently, our study focused on short-term interactions with AI, which may not fully reflect the complexities and dynamics of real-world AI usage over longer periods. Longitudinal studies could provide deeper insights into how sustained engagement with AI writing aids might influence individuals' locus of control and professional identity over time. Additionally, investigating how different interface designs might mitigate the negative effects on agency for unemployed users would be valuable for developing more inclusive AI systems.

\section{Conclusion}
How deeply does AI reach into our sense of self, and how can we design systems that carefully consider this reach? With 462 participants, the current study observed the different shifts in the locus of control between employed and unemployed individuals, reflecting AI's far-reaching influence on self and identity. These findings reveal a critical insight: AI systems have the power to influence not just what we do, but how we see ourselves. Beyond shaping our actions, knowledge, and social interactions, AI reaches deep into the core of who we are. 

\section*{Acknowledgments}
This material is based upon work supported by the National Science Foundation under Grant No. IIS-2212396. We thank Mor Naaman, Interplay Research Studio, and the Social Technologies research group for their feedback on research design and ideation.

\bibliography{anthology}

\end{document}